\documentclass{phostproc}

\usepackage{lipsum} %%% REMOVE THIS

\title{K2 Observations of Galactic RRd Stars}
\author{Pawe{\l} Moskalik,$^{1}$
        James M. Nemec,$^{2}$
        L\'aszl\'o Moln\'ar,$^{3,4}$
        Emese Plachy,$^{3,4}$
        R\'obert Szab\'o,$^{3,4}$
        Katrien Kolenberg$^{5,6}$
        }

\affiliation{$^{1}$ Nicolaus Copernicus Astronomical Center, Warsaw, Poland \\
             $^{2}$ Department of Physics \& Astronomy, Camosun College, Victoria, Canada \\
             $^{3}$ Konkoly Observatory, MTA CSFK, Budapest, Hungary \\
             $^{4}$ MTA CSFK Lend\"ulet Near-Field Cosmology Research Group, Budapest, Hungary \\
             $^{5}$ Institute of Astronomy, KU Leuven, Heverlee, Belgium \\
             $^{6}$ Physics Department, University of Antwerp, Antwerp, Belgium
             }

\shorttitle{K2 Observations of Galactic RRd Stars}
\shortauthors{Pawe{\l} Moskalik \textit{et al.}}
\abs{We have analysed high-precision photometry of 77 double-mode
RR~Lyrae (RRd) stars observed by NASA's {\it Kepler-K2} mission
during Campaigns 1-18. Among those stars, we have identified several
low-period ratio variables, belonging either to a subclass of
"anomalous" RRd stars \citep{anomRRd} or to a separate
shorter-period subclass recently identified by \cite{Prudil2017}. A
non-radial mode with the period ratio of $P_{\rm x}/P_{\rm 1} \sim
0.615$ has been detected in most RRd stars of our sample. In
majority of these variables, at least one subharmonic of the
non-radial frequency is also present. Our findings indicate that
excitation of the non-radial mode is a common property of the RRd
stars. The same non-radial pulsation is also commonly detected in
the RRc stars. We show that properties of this puzzling mode are
essentially the same in both groups of RR Lyrae-type pulsators.}

\begin{document}

\maketitle

\section{Introduction}

RRd stars are a subclass of the RR Lyrae variables, in which two
radial modes, the fundamental and the first overtone, are
simultaneously excited. Such variables are rare and not a single one
has been found the original {\it Kepler} field \citep{Nemec2013}.
Only with the {\it Kepler-K2} mission a high precision space
photometry has finally been collected for a large sample of RRd
stars. We are now in position to study their pulsation in great
detail.

In this report we discuss 77 RRd variables observed during {\it K2}
Campaigns 1-18. For most of these objects we have derived accurate
photometry either with the EAP pipeline \citep{EAP} or with the PyKE
pipeline \citep{PYKE}. For the remaining ones, we have used the
standard {\it Kepler} PDCSap photometry provided by the {\it K2}
archive. The latter is generally of somewhat lower quality. For all
the stars we have performed detailed frequency analysis, using a
standard Fourier Transform, followed by a consecutive frequency
prewhitening. We note that four stars of our sample have already
been studied before by other authors \citep{Molnar2015, Kurtz2016,
Plachy2017}.

A paper presenting full results of our analysis is being readied for
publication \citep{Nemec2019}.

\section{RRd variables in K2 fields}

\begin{figure}
\vskip 3.7truecm
    \centering
    \includegraphics[width=0.95\linewidth]{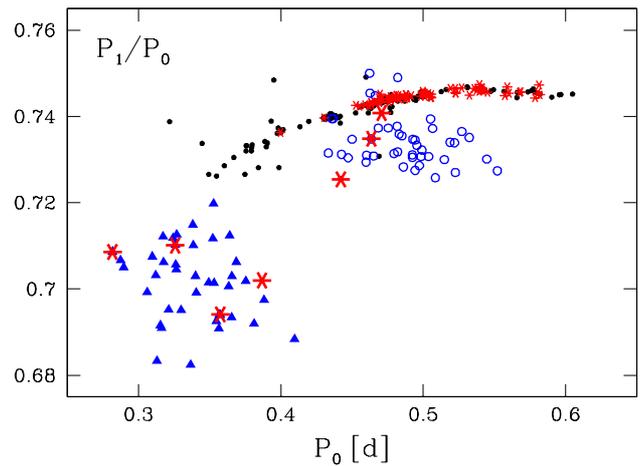}
    \caption{Petersen diagram for radial modes of RRd stars. The K2
             RRd stars are plotted with red asterisks. "Classical"
             RRd stars, anomalous RRd stars and "Prudil's stars" are
             displayed with black dots, open blue circles and blue
             triangles, respectively.}
    \label{fig1}
\end{figure}

In Fig.\thinspace\ref{fig1} we present the Petersen diagram (the
period ratio vs. period diagram) for the {\it K2} RRd stars (plotted
with red asterisks). For comparison, we also display the "classical"
RRd stars of the Galactic bulge \citep{BulgeRRd}, the anomalous RRd
stars collected from several stellar systems \citep{Smolec2015, M3,
anomRRd, NGC6362} and the so-called "Prudil's stars"
\citep{Prudil2017}.\footnote{ For full discussion of different
subclasses of the RRd pulsators the reader is referred to
\cite{Revol}.}

Most of the RRd variables of the {\it K2} sample are typical
double-mode RR Lyrae pulsators and fall on a tight progression
defined by the classical RRd stars. There are several objects,
though (marked with large red symbols), which behave differently.
Three {\it K2} variables are placed firmly among anomalous RRd
pulsators. In two of them pulsation modes are modulated
\citep{Smolec2015, Plachy2017}. This is a very frequent phenomenon
in anomalous RRd stars, but is not observed in the classical RRd
pulsators. Four other {\it K2} variables with low period ratios and
short periods belong to a subclass of the "Prudil's stars". In all
seven exceptional objects, pulsations are strongly dominated by the
radial fundamental mode $A_{\rm 0}/A_{\rm 1} = 1.4-7.6$, which is
another difference between them and the classical RRd pulsators.

\section{Secondary modes}

\begin{figure*}[ht]
    \centering

    \includegraphics[width=0.85\textwidth]{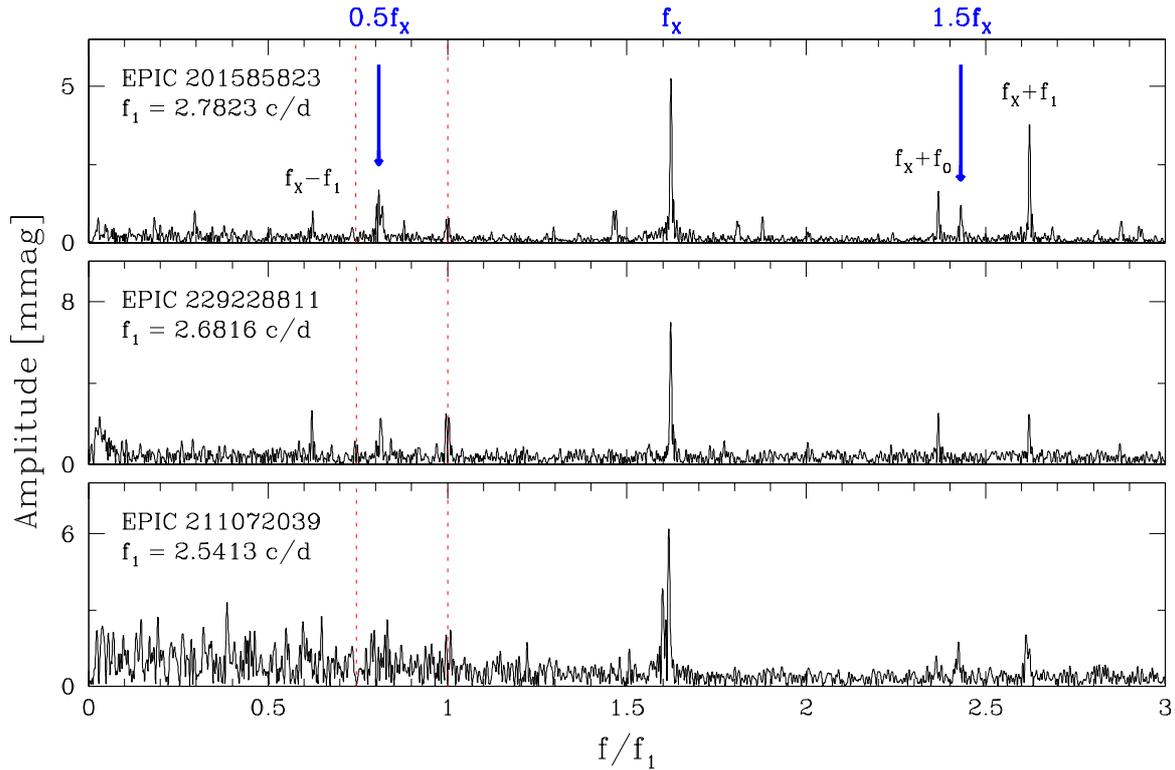}
    \caption{Fourier frequency spectra of 3 RRd stars, after
             prewhitening the data of all the signal related to the
             radial modes. Red vertical dashed lines mark
             frequencies of the removed radial modes.}

    \label{fig2}
\end{figure*}

For all RRd stars of our sample we have conducted a search for
secondary low-amplitude pulsations. After prewhitening the data of
the two radial modes, their harmonics and their combination
frequencies, we have detected in many {\it K2} RRd variables an
additional variability with frequency, $f_{\rm x}$. Linear
combinations of $f_{\rm x}$ with frequencies of the radial modes are
also usually present. The period ratio of this secondary mode to the
first radial overtone, $P_{\rm x}/P_{\rm 1}$, is always close to
0.615 and its amplitude is always very low and rarely exceeds
10mmag. This is the same period ratio which is commonly observed in
the first overtone RR Lyrae (RRc) stars \citep{KepRRc, BulgeRRc2}.
This period ratio cannot be explained with any two radial modes,
implying that frequency $f_{\rm x}$ must be associated with a
nonradial mode of pulsation. In many RRd stars a significant signal
has also been found at $\sim\! 0.5f_{\rm x}$ or $\sim\! 1.5f_{\rm
x}$, or both. These subharmonics always appear together with the
parent mode, $f_{\rm x}$, and never alone. The presence of the
subharmonic (half-integer) frequencies is another similarity to the
RRc variables. Interestingly, the $f_{\rm x}$ mode has been detected
only in the classical RRd variables. It has not been detected in any
of the "Prudil's stars" or the anomalous RRd stars.

In Fig.\thinspace\ref{fig2} we display Fourier frequency spectra for
3 classical RRd stars. The spectra are computed after prewhitening
the data of all the signal related to the radial modes. The
frequency axis is normalized by the frequency of the first radial
overtone, $f_{\rm 1}$. Such a normalization is adopted to better
show how similar the frequency patterns in different RRd stars are.
In all 3 variables of Fig.\thinspace\ref{fig2} the $f_{\rm x}$ mode
is unambiguously detected. Its subharmonics are also clearly seen,
although only in two variables both subharmonics are present. The
Fourier peak at $f_{\rm x}$ is sometimes visibly broadened or split,
indicating that the mode is not stationary. A detailed
time-dependent analysis reveals that both the amplitude and the
phase of $f_{\rm x}$ mode are strongly variable. This is in contrast
to the properties of the radial modes, which are usually almost
perfectly stable.

It is interesting to establish how often the $f_{\rm x}$ mode is
excited in the RRd stars. Because this mode is always very weak, we
have to use the best data available to get this estimate. Therefore,
we limit our sample to only those stars for which EAP or PyKE
photometry is available. Furthermore, we limit the sample to the
classical RRd stars, the only subclass in which $f_{\rm x}$ mode is
detected. Such restricted subsample consists of 57 variables. Among
those objects, the $f_{\rm x}$ mode is clearly detected in 43 RRd
variables, which is 75\% of the sample. Subharmonics of $f_{\rm x}$
have lower amplitude and are even more difficult to detect. Yet,
they are found in 34 RRd variables. This constitutes 60\% of the RRd
sample and 79\% of the stars in which $f_{\rm x}$ is present.

We can restrict the sample even further by accepting only stars with
EAP photometry. The EAP data are somewhat better than PyKE data,
especially when the field is crowded. The new subsample is limited
to only 40 RRd stars, but this is the highest quality subsample we
can select. The incidence rate of detecting the $f_{\rm x}$
frequency increases now to 85\% (34 stars) and of detecting its
subharmonic to 68\% (27 stars). Clearly, the better the data the
higher the probability of finding the $f_{\rm x}$ mode. This
indicates that we have probably not yet reached the true incidence
rates and the numbers given above should be treated only as lower
limits. We conclude that excitation of the $f_{\rm x}$ mode and its
subharmonics is very common in the RRd stars. The mode appears in
the majority and possibly in all of the classical RRd stars. In most
cases, the mode is accompanied by at least one subharmonic.

\section{Similarity of RRc and RRd stars}

\begin{figure}
\vskip 3.7truecm
    \centering
    \includegraphics[width=0.95\linewidth]{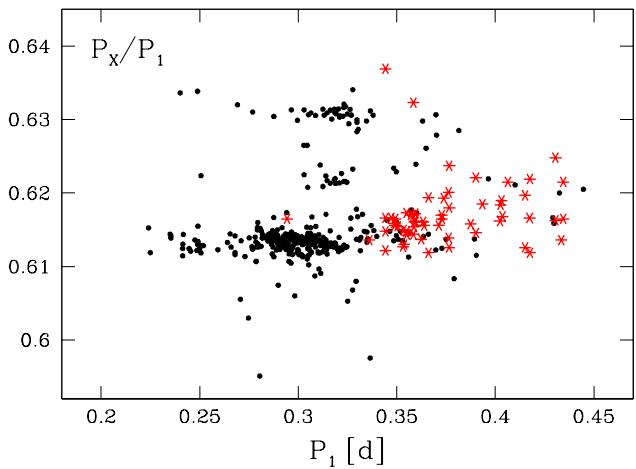}
    \caption{Petersen diagram for secondary modes of RRc and RRd stars.
             Period ratios are plotted vs. first overtone period,
             $P_1$. The RRc stars of the Galactic Bulge
             \citep{BulgeRRc1, BulgeRRc2} are displayed with black
             dots. The RRd variables of the {\it K2} sample are
             plotted with red asterisks.}
    \label{fig3}
\end{figure}

Secondary modes with period ratio of $P_{\rm x}/P_{\rm 1} = 0.60 -
0.64$ have been detected in nearly 300 RRc variables \citep{Revol}.
On the Petersen diagram the RRc stars form three well-separated
horizontal sequences, which are centered on $P_{\rm x}/P_{\rm 1} =
0.613$, $0.632$ and $0.631$ (Fig.\thinspace\ref{fig3}). The RRd
stars of our sample fit into this pattern very nicely. Vast majority
of them are placed on the lowest of the three sequences. This is to
be expected, because the lowest sequence is the most populated among
the three. The properties of the $f_{\rm x}$ mode are also the same
in both groups of variables. In both RRc and RRd stars, the $f_{\rm
x}$ modes:

\begin{itemize}
\item have similar period ratio with the first radial overtone;
\item are usually accompanied by subharmonics (half-integer
      frequencies) at $\sim\! 0.5f_{\rm x}$ and/or at $\sim\!
      1.5f_{\rm x}$.
\item are detected in almost every star for which high precision space
      photometry is available (see \cite{KepRRc} for discussion of
      the incidence rate in RRc stars).
\item have amplitudes in the same range;
\item are nonstationary, while dominant radial modes are usually almost
      perfectly stable;
\end{itemize}

\noindent The similarity of the $f_{\rm x}$ mode in the RRd and the
RRc variables is really striking. Clearly, in both subgroups of
RR~Lyrae pulsators we observe the same phenomenon. Consequently, for
both RRd and RRc stars a common explanation has to be found for
$f_{\rm x}$. Because of the observed period ratio, this signal
cannot be identified with any radial mode. Thus, the $f_{\rm x}$
frequency must be associated with a nonradial mode of pulsation. A
promising interpretation has recently been put forward by \cite{WD}.
In the proposed scenario, the $f_{\rm x}$ frequency corresponds to
the first harmonic of the excited mode of $\ell=8$ or $\ell=9$.

\section*{Acknowledgments}
This research was supported in part by the National Science Center,
Poland, through grant No DEC-2015/17/B/ST9/03421 (PM). It was also
supported by the NKFIH grants K-115709, PD-121203 and PD-116175 (LM,
EP, RSz), the Ja\'nos Bolyai Research Scholarship, the LP2014-17
(LM, EP) and LP2018-7 (LM, EP, RSz) Lend\"ulet Programs of the HAS
and by the FWO-PAS Poland-Belgium scientific cooperation grant
VS.091.16N (PM, KK).

\bibliographystyle{phostproc}
\bibliography{pam.bib}

\end{document}